\documentclass[aps,prx,reprint,superscriptaddress,longbibliography]{revtex4-2}

\usepackage{graphicx}
\usepackage{amsmath,amssymb}
\usepackage{hyperref}
\usepackage{booktabs}
\usepackage{tabularx}

\begin{document}


\title{Recoverable Quantum Computation: An Information-Centric Paradigm for Quantum Computing with Errors}

\author{Shengwang Du}
\email{dusw@purdue.edu}
\affiliation{Elmore Family School of Electrical and Computer Engineering, Purdue University, West Lafayette, Indiana 47907, USA}
\affiliation{Department of Physics and Astronomy, Purdue University, West Lafayette, Indiana 47907, USA}
\affiliation{Purdue Quantum Science and Engineering Institute, Purdue University, West Lafayette, Indiana 47907, USA}

\begin{abstract}

Quantum computing promises transformative advances in computation, communication, sensing, and machine learning. Yet the realization of large-scale fault-tolerant quantum computers remains hindered by the enormous overhead required for quantum error correction. This challenge raises a fundamental question: \emph{Must useful quantum computing wait until fully fault-tolerant quantum hardware becomes available?}  In this Perspective, we propose \emph{Recoverable Quantum Computation} (RQC), an information-centric paradigm for quantum computing with errors. Rather than requiring faithful preservation of the complete quantum state, RQC focuses on preserving the computational information required to accomplish a given task. A quantum computation is considered recoverable if the desired computational information can be extracted from noisy quantum outputs with an overhead that preserves quantum advantage relative to the best known classical method. We introduce recoverability as an operational principle for evaluating noisy quantum computations and propose practical metrics based on recovery overhead and recoverability efficiency. We illustrate the framework using quantum Fourier transform period estimation on IBM quantum hardware and a conceptual example from quantum machine learning, demonstrating that useful computational information may remain recoverable despite significant physical errors. Building on these examples, we propose a preliminary classification of quantum applications according to their expected recoverability and outline a research roadmap toward a predictive theory of recoverability. RQC is intended not as an alternative to fault-tolerant quantum computing, but as a complementary paradigm for understanding and evaluating useful quantum computation in the broad intermediate regime between today's noisy quantum processors and tomorrow's fault-tolerant quantum computers.

\end{abstract}

\maketitle

\section{Introduction} \label{sec:Introduction}

\begin{figure*}[t]
\centering
\includegraphics[width=\textwidth]{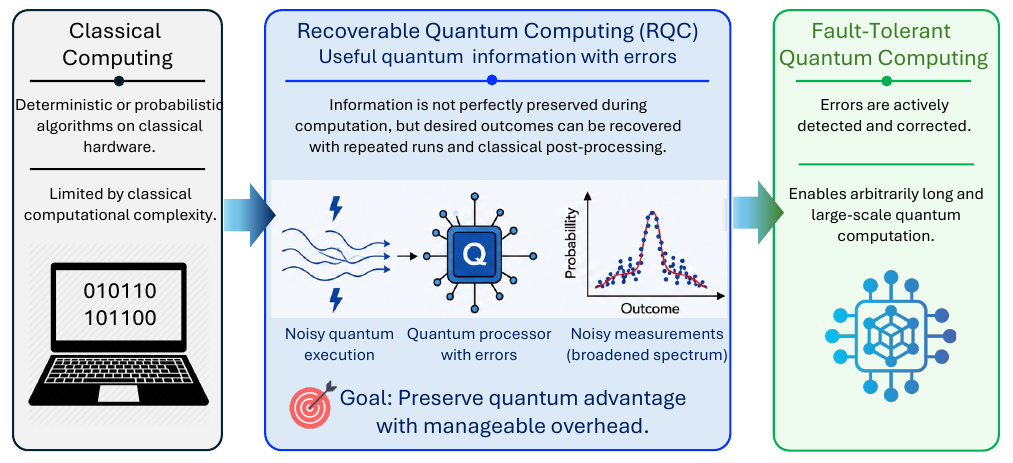}
\caption{Conceptual relationship among classical computing, recoverable quantum computation (RQC), and fault-tolerant quantum computing (FTQC). Classical computing is limited by classical computational complexity. FTQC achieves arbitrarily accurate quantum computation through active error detection and correction. RQC occupies an intermediate regime in which useful computational outcomes can be extracted from noisy quantum processors through repeated executions and classical post-processing while preserving quantum advantage with manageable overhead.}
\label{fig:RQC_paradigm}
\end{figure*}

Quantum computing promises transformative advances in computation, communication, sensing, and machine learning~\cite{Feynman1982,Deutsch1985,NielsenChuang}. Over the past decades, remarkable progress has been achieved across a broad range of hardware platforms, including superconducting circuits, trapped ions, neutral atoms, photonic systems, and semiconductor quantum devices~\cite{npjQI-Matthias,IonQC-Sage,PhysicsToday-Saffman,PhysRevLett.104.010503,Arute2019,Zhong2020,Acharya2023,Bluvstein2024}. Quantum processors containing tens to thousands of physical qubits have now been demonstrated, while landmark advances in quantum advantage, quantum simulation, variational algorithms, and quantum machine learning have fueled growing expectations for practical quantum technologies~\cite{Arute2019,Preskill2018,Peruzzo2014,Biamonte2017}. Despite this remarkable progress, a fundamental challenge remains. Quantum systems are inherently susceptible to errors arising from decoherence, imperfect control, environmental coupling, crosstalk, state-preparation imperfections, and measurement noise~\cite{NielsenChuang,LidarBrun}. As quantum processors scale to larger systems and deeper circuits, these errors accumulate and eventually degrade computational performance.

The prevailing solution is fault-tolerant quantum computing (FTQC), in which logical quantum information is protected by encoding it into many physical qubits through quantum error correction~\cite{Shor1995,Steane1996,Gottesman1998}. Supported by the quantum threshold theorem~\cite{AharonovBenOr1997,Knill1998,Aliferis2006}, this framework provides a rigorous path toward arbitrarily accurate quantum computation. However, this capability comes at a substantial cost. Depending on the hardware platform and physical error rates, a single logical qubit may require hundreds or even thousands of physical qubits, while a practical fault-tolerant quantum computer capable of solving large-scale problems may ultimately requires millions of physical qubits~\cite{Fowler2012}, which significantly increases the challenges in connectivity~\cite{SQGPU}. Although impressive progress toward logical-qubit demonstrations continues~\cite{Acharya2023,Quantinuum2024}, the hardware overhead required for fault tolerance remains one of the greatest obstacles to practical quantum computing.

This situation naturally raises a fundamental question:
\begin{quote}
\emph{Must useful quantum computing wait for fully fault-tolerant quantum hardware?}
\end{quote}
Much of the current discussion implicitly equates useful quantum computation with FTQC. Under this viewpoint, today's noisy quantum processors primarily serve as intermediate platforms for demonstrating the technologies required for future fault-tolerant systems. Although this perspective is essential for scalable quantum computation, it may be unnecessarily restrictive for many practically useful applications.

The history of information science suggests a broader perspective~\cite{Landauer1991}. Reliable communication routinely occurs over noisy channels because the transmitted information can be recovered despite imperfections~\cite{Shannon1948}. Likewise, experimental measurements in physics are almost never noise free, yet meaningful information can often be extracted through repeated measurements, statistical averaging, parameter estimation, and inference. In both cases, the essential question is not whether the underlying process is perfectly error free, but whether the desired information remains recoverable with acceptable overhead.

Quantum computing may also admit such an intermediate regime. Most quantum algorithms do not require complete reconstruction of the final quantum state. Instead, they seek specific computational information encoded in measurement outcomes, such as a frequency, a phase, a classification label, an optimized parameter, or a physical observable. This task-specific information often represents only a small fraction of the exponentially large information contained in the full quantum state. Consequently, preserving the complete quantum state may represent a substantially stronger requirement than preserving the information needed to solve the computational task.

Motivated by this observation, we propose \emph{Recoverable Quantum Computation} (RQC), an information-centric paradigm for quantum computing with errors. Rather than demanding faithful preservation of the complete quantum state, RQC focuses on preserving the computational information required to accomplish a given task. A quantum computation is considered recoverable if the desired computational information can be extracted from noisy quantum outputs with an overhead that preserves quantum advantage relative to the classical method.

RQC is related to, but fundamentally distinct from, existing efforts in noisy intermediate-scale quantum (NISQ) computing and quantum error mitigation~\cite{Preskill2018,Temme2017,Endo2018,Cai2021,Huggins2021}. NISQ computing describes a hardware regime characterized by limited qubit counts and imperfect operations, whereas error mitigation develops techniques for reducing the effects of hardware noise on measured observables. RQC addresses a different question: under what conditions does noisy quantum computation remain useful? Rather than focusing primarily on hardware fidelity, it asks whether the computational information required by a given task remains recoverable despite imperfect quantum execution.

Figure~\ref{fig:RQC_paradigm} illustrates the conceptual relationship among classical computing, RQC, and FTQC. Classical computation is fundamentally limited by classical computational complexity, whereas FTQC seeks arbitrarily accurate quantum information processing through active quantum error correction. RQC occupies the broad intermediate regime in which useful computational information remains accessible despite noisy quantum execution. In this regime, repeated quantum executions together with classical post-processing can compensate for hardware imperfections while preserving practical quantum advantage.

We anticipate that RQC could provide a useful conceptual framework for understanding the broad landscape between today's noisy quantum processors and tomorrow's fault-tolerant quantum computers. By shifting the focus from preserving quantum states to preserving computational information, RQC offers a complementary perspective on when useful quantum advantage can be achieved before fully fault-tolerant quantum hardware becomes available.

It is important to emphasize that RQC does not propose new quantum algorithms or new computational objectives. Quantum algorithms have always been designed to solve application-specific problems such as estimating a physical parameter, identifying a solution, or producing a classification decision. The novelty of RQC lies instead in the criterion used to evaluate noisy quantum computation. Rather than asking whether the underlying quantum state has been faithfully preserved, RQC asks whether the computational objective itself remains recoverable with an overhead that preserves quantum advantage. In this sense, RQC is not an alternative class of quantum algorithms, but a complementary framework for understanding when noisy quantum computation remains practically useful. The central contribution of this Perspective is therefore not a new quantum algorithm, but a complementary framework for evaluating the practical usefulness of noisy quantum computation.

The remainder of this Perspective is organized as follows. Section~\ref{sec:Why} discusses why fault tolerance is not the only possible path toward useful quantum computation. Section~\ref{sec:RQC} introduces RQC and develops a hierarchy of recoverability. Section~~\ref{sec:Metrics} presents operational metrics for quantifying recoverability. Sections~\ref{sec:QFT} and \ref{sec:QML} illustrate the concept through quantum Fourier transform period estimation and quantum machine learning, respectively. Section~\ref{sec:Classification} proposes a preliminary classification of quantum applications according to their expected recoverability. Section~\ref{sec:Roadmap} outlines a research roadmap toward a predictive theory of recoverability, and Section~\ref{sec:Conclusions} concludes with a broader perspective on quantum computation with errors.

\section{Why Fault Tolerance Is Not the Only Path} \label{sec:Why}

FTQC represents one of the greatest achievements in quantum information science. The development of quantum error correction and the quantum threshold theorem established that, in principle, arbitrarily long quantum computations can be performed reliably despite the presence of physical errors~\cite{Shor1995,Steane1996,Gottesman1998,AharonovBenOr1997,Knill1998,Aliferis2006}. Consequently, fault tolerance has become the dominant paradigm for scalable quantum computation and is widely regarded as the ultimate destination of quantum computing technology.

From this perspective, useful quantum computation appears inseparable from FTQC. If errors accumulate during a computation, quantum information is progressively degraded until computational correctness can no longer be guaranteed. As a result, much of today's hardware development is devoted to reducing physical error rates and building quantum processors capable of fault-tolerant operation.

Although this viewpoint is both natural and compelling, it rests on an implicit assumption: that useful quantum computation requires faithful preservation of the complete quantum state throughout the computation. This assumption is entirely appropriate from the perspective of quantum error correction, whose objective is to protect the logical quantum state against arbitrary physical errors. Many practical quantum algorithms, however, ultimately require only the computational information encoded in the final measurement outcomes rather than complete knowledge of the underlying quantum state.

We believe this distinction is fundamental. A quantum state generally contains an exponentially large amount of information, whereas the information required to solve a particular computational task is often much smaller. For example, a quantum Fourier transform is used to estimate a frequency or phase, a quantum classifier outputs a decision label, and a quantum optimization algorithm seeks an optimal parameter or solution. In each case, the desired result represents only a small fraction of the information encoded in the underlying quantum state.

A useful analogy can be drawn from classical communication theory. Communication channels are inherently noisy, yet reliable communication remains possible because the transmitted information can be recovered despite imperfections~\cite{Shannon1948}. The objective is not to eliminate all noise from the communication channel, but rather to ensure that the desired information remains recoverable with acceptable overhead. Experimental physics provides another familiar example. Photon shot noise, thermal fluctuations, detector imperfections, and environmental disturbances routinely affect measurements. Nevertheless, meaningful physical information can often be extracted through repeated measurements, statistical averaging, parameter estimation, and statistical inference. Experimental physicists rarely ask whether a measurement is perfectly noise free; instead, they ask whether the quantity of interest can be determined with sufficient confidence.

Quantum computing may admit a similar intermediate regime. Between today's noisy quantum processors and tomorrow's fully fault-tolerant quantum computers lies a broad region in which useful computational information may remain recoverable despite imperfect quantum execution. In this regime, hardware errors degrade the quality of the quantum state without necessarily destroying the information relevant to the computational task. Repeated quantum executions, statistical inference, and classical post-processing may therefore compensate for imperfect hardware while preserving a computational advantage over the best known classical methods.

Recent developments in NISQ computing~\cite{Preskill2018}, quantum error mitigation~\cite{Temme2017,Endo2018,Cai2021,Huggins2021}, variational quantum algorithms~\cite{Peruzzo2014,Cerezo2021}, quantum machine learning~\cite{Biamonte2017,Havlicek2019,Schuld2019}, and quantum sensing have already hinted at this possibility. These developments suggest that useful quantum computation may be achievable under less restrictive conditions than those required for full fault tolerance. Nevertheless, these approaches have largely been developed as techniques for improving the performance of noisy quantum processors rather than as a broader framework for understanding when useful computational information remains recoverable.

This observation motivates the central question addressed in this Perspective:
\begin{quote}
\emph{Under what conditions can useful computational information remain recoverable even when the underlying quantum computation is imperfect?}
\end{quote}
RQC addresses this question by shifting the focus from preserving the complete quantum state to preserving the computational information required by the task. From this perspective, the relevant criterion is not whether the quantum evolution is perfectly faithful, but whether the desired computational information can be recovered with an overhead that preserves quantum advantage.

This viewpoint naturally leads to three complementary computational regimes:
\begin{enumerate}
\item \textbf{Classical computation}, in which performance is fundamentally limited by classical computational complexity.

\item \textbf{Recoverable quantum computation (RQC)}, in which useful computational information remains recoverable despite noisy quantum execution.

\item \textbf{Fault-tolerant quantum computation (FTQC)}, in which active quantum error correction enables arbitrarily accurate quantum information processing.
\end{enumerate}
FTQC remains the ultimate solution for scalable quantum information processing. RQC neither replaces fault tolerance nor diminishes the importance of quantum error correction. Rather, it identifies a potentially broad intermediate regime in which useful quantum computation may be achieved before fully fault-tolerant hardware becomes available.

The existence, boundaries, and practical significance of this intermediate regime remain open questions. Addressing these questions requires a framework for defining recoverability, identifying the computational information relevant to different quantum algorithms, and quantifying the overhead required to recover that information. Developing such a framework is the objective of the following sections.

\section{Recoverable Quantum Computation (RQC)} \label{sec:RQC}

The discussion in the previous section suggests that useful quantum computation does not necessarily require faithful preservation of the complete quantum state. Instead, many quantum algorithms ultimately require only the computational information extracted from the final measurement outcomes. This observation motivates a shift in perspective. Rather than asking whether a quantum computation faithfully preserves the entire quantum state, RQC asks whether the computational information required to accomplish a given task remains recoverable despite imperfect quantum execution.

Throughout this Perspective, we use the term \emph{computational information} to denote the minimum task-specific information required to successfully complete a computational objective. Depending on the application, this information may consist of a physical parameter estimated by a quantum Fourier transform, a classification label produced by a quantum machine-learning algorithm, an optimized variable obtained through variational optimization, or a physical observable measured in quantum sensing. Importantly, computational information should be distinguished from the complete quantum state. While an $n$-qubit quantum state generally contains exponentially many probability amplitudes, many practical quantum algorithms ultimately require only a small fraction of this information to accomplish their intended task. Identifying rigorous mathematical definitions of computational information for general quantum algorithms remains an open problem and represents an important direction for future research. From this perspective, computational information is an application-dependent concept rather than an intrinsic property of a quantum state.

Traditional FTQC seeks to preserve the entire logical quantum state throughout a computation. RQC adopts a different evaluation criterion. Rather than evaluating noisy quantum computation primarily through preservation of the complete quantum state, it evaluates whether the computational information required by the task remains recoverable. From this perspective, errors become significant not simply because they reduce quantum-state fidelity, but because they may prevent recovery of the information needed to complete the computational task.

\subsection{A Hierarchy of Recoverability}

This distinction motivates three complementary notions of recoverability.

\textbf{State Recoverability.}
The strongest notion of recoverability concerns the quantum state itself. FTQC seeks to preserve the encoded logical quantum state throughout the computation by means of active quantum error correction. This objective forms the foundation of scalable fault-tolerant quantum computation.

\textbf{Computational Information Recoverability.}
A weaker, yet considerably more general, notion concerns the recoverability of the computational information required by a specific algorithm. Many quantum algorithms ultimately seek only a frequency, a phase, a classification label, an optimized parameter, or another task-specific quantity. In such cases, faithful reconstruction of the complete quantum state is unnecessary provided that the desired computational information remains recoverable.

\textbf{Advantage Recoverability.}
The most practically relevant notion concerns the preservation of quantum advantage. Even if repeated quantum executions, statistical inference, or classical post-processing are required, a quantum algorithm remains useful as long as its total computational cost remains lower than that of the best known classical method.

Accordingly, the RQC framework proposed here focuses primarily on not with state recoverability, but with computational-information recoverability and, ultimately, the preservation of quantum advantage.

\subsection{Definition of Recoverable Quantum Computation}

Motivated by the preceding discussion, we propose the following working definition.
\medskip

\noindent
\textbf{Recoverable Quantum Computation (RQC).}
\emph{A quantum computation is recoverable if the computational information required to accomplish a given task can be extracted from noisy quantum outputs with an overhead that preserves quantum advantage relative to the classical method.}

\medskip

Several aspects of this definition merit emphasis. First, recoverability is inherently application dependent. Different quantum algorithms encode computational information in different ways and therefore exhibit different sensitivities to hardware noise. Second, recoverability is fundamentally distinct from quantum-state fidelity. A computation with relatively low state fidelity may nevertheless preserve the computational information required by the application. Conversely, a computation exhibiting high state fidelity may still fail to provide useful results if the desired computational information cannot be recovered efficiently. Third, recoverability does not require active quantum error correction. Repeated quantum executions, statistical estimation, classical inference, machine learning, or other post-processing techniques may all contribute to recovering useful computational information.

\subsection{Computational Information as a Resource}

The RQC framework proposed here motivates a different perspective on computational resources. In conventional FTQC, the principal resource being protected is the logical quantum state. Considerable physical resources are devoted to maintaining high quantum-state fidelity throughout the computation. RQC instead treats computational information as the primary resource. Rather than correcting every physical error, one allows imperfect quantum execution and compensates through repeated quantum measurements combined with classical post-processing. The tradeoff therefore shifts from physical-qubit overhead to recovery overhead. Conceptually, this distinction resembles the difference between error correction and statistical estimation. Error correction seeks to prevent information loss, whereas statistical estimation seeks to recover useful information despite imperfect measurements. RQC adopts the latter perspective for quantum computation.

\subsection{Relation to Existing Paradigms}

RQC is related to, but fundamentally distinct from, several existing paradigms. FTQC seeks to preserve the complete logical quantum state through active quantum error correction. Quantum error mitigation seeks to suppress or compensate for the effects of hardware noise on measured observables. NISQ computing describes a hardware regime characterized by limited qubit counts and imperfect operations~\cite{Preskill2018}. RQC instead proposes a criterion for determining when noisy quantum computation remains useful. It asks whether the computational information required by a quantum algorithm remains recoverable despite imperfect quantum execution. From this perspective, FTQC, quantum error mitigation, and RQC emphasize three complementary aspects of noisy quantum computation: state preservation, error suppression, and information preservation, respectively.

The distinction between RQC and quantum error mitigation deserves particular emphasis. Quantum error mitigation comprises a family of techniques that estimate or approximate ideal quantum outcomes by compensating for hardware errors through methods such as probabilistic error cancellation, zero-noise extrapolation, virtual distillation, symmetry verification, or other classical post-processing strategies. RQC addresses a broader and different question: under what conditions does a noisy quantum computation remain practically useful? From the perspective of RQC, quantum error mitigation may serve as one possible mechanism for improving recoverability, but it is neither required nor sufficient to define recoverability itself. A quantum computation may remain recoverable without explicit error-mitigation techniques. Conversely, even highly successful error mitigation does not by itself guarantee recoverability if the total recovery overhead ultimately eliminates the computational advantage over the best known classical method. In this sense, recoverability is an application-level concept, whereas error mitigation is a hardware- and algorithm-level methodology.

\subsection{Toward Quantifying Recoverability}

The concepts introduced above remain qualitative. Transforming RQC into a predictive theoretical framework would requires quantitative measures of recoverability. The central question is no longer simply how accurately a quantum processor preserves a quantum state, but how efficiently the desired computational information can be recovered from noisy quantum outputs. If the additional resources required for recovery remain modest, quantum advantage may survive despite substantial hardware errors. Conversely, if the recovery overhead grows too rapidly with problem size, the practical advantage of quantum computation eventually disappears. These observations motivate the recoverability metrics introduced in the next section. The subsequent examples of quantum Fourier transform and quantum machine learning illustrate two representative forms of computational-information recovery: parameter recovery and decision recovery.

\section{Recoverability Metrics}\label{sec:Metrics}

The previous section introduced RQC as an information-centric paradigm in which the objective is to preserve the computational information required by a quantum algorithm rather than the complete quantum state. To determine whether such information remains useful in practice, recoverability must be quantified through operational metrics. The goal of this section is therefore not to develop a complete mathematical theory of recoverability, but rather to establish a practical framework for measuring how efficiently computational information can be recovered from noisy quantum outputs. The central question is straightforward: \emph{How much additional effort is required to recover the desired computational information from a noisy quantum computation?} We refer to this additional effort as the \emph{recovery overhead}.

\subsection{Recovery Overhead}

Consider a quantum algorithm that, under ideal conditions, produces the desired computational information in a single execution. On a realistic noisy quantum processor, hardware imperfections broaden output distributions, reduce confidence in measurement outcomes, and increase uncertainty in the recovered quantities. Consequently, repeated quantum executions may be required before the desired computational information can be extracted with a specified level of confidence.

Let $N_{\mathrm{rec}}$ denote the number of quantum executions required to recover the desired computational information with a prescribed confidence level. We refer to $N_{\mathrm{rec}}$ as the \emph{recovery overhead}. For an ideal quantum computer, $N_{\mathrm{rec}}$ approaches unity. On noisy quantum hardware, however, $N_{\mathrm{rec}}$ generally increases as hardware noise and circuit errors accumulate. From the perspective of RQC, the crucial issue is therefore not whether repeated executions are required, but rather how the recovery overhead scales with problem size.

\subsection{Recoverability and Scaling}

The practical usefulness may depend critically on the scaling behavior of the recovery overhead. Suppose that a problem of size $n$ requires
\begin{equation}
N_{\mathrm{rec}}(n)=\mathrm{poly}(n),
\end{equation}
quantum executions to recover the desired computational information. In this regime, the recovery overhead grows only polynomially with problem size, allowing the quantum algorithm to retain a practical advantage over the best known classical approach. By contrast, if
\begin{equation}
N_{\mathrm{rec}}(n)\sim e^{\alpha n},
\end{equation}
where $\alpha>0$, the recovery overhead grows exponentially. In this case, the cost of recovering the desired computational information eventually overwhelms the original quantum advantage.

This observation naturally suggests two qualitative computational regimes:
\begin{itemize}
\item \textbf{Recoverable regime}: the recovery overhead scales polynomially with problem size.

\item \textbf{Unrecoverable regime}: the recovery overhead scales exponentially with problem size.
\end{itemize}
Although intentionally simple, this distinction captures the central intuition underlying RQC: the practical usefulness of noisy quantum computation depends not only on hardware fidelity, but also on how efficiently the desired computational information can be recovered as the problem size increases.

\subsection{Recoverability Efficiency}

Ultimately, recoverability should be evaluated relative to the best available classical algorithm. Let $T_Q$ denote the runtime of a single quantum execution and $T_C$ the runtime of the best known classical algorithm. The total runtime required to recover the desired computational information is approximately
\begin{equation}
T_{\mathrm{RQC}}\approx N_{\mathrm{rec}}T_Q.
\end{equation}
We therefore define the recoverability efficiency as
\begin{equation}
\eta
=
\frac{T_C}{T_{\mathrm{RQC}}}
=
\frac{T_C}{N_{\mathrm{rec}}T_Q},
\label{eq:eta}
\end{equation}
which quantifies the practical usefulness of a noisy quantum computation rather than the fidelity of its underlying quantum state. A value of $\eta>1$ indicates that the recoverable quantum computation retains an advantage over the best known classical method, whereas $\eta<1$ indicates that the recovery overhead has eliminated the practical quantum advantage.

Equation~(\ref{eq:eta}) should be regarded primarily as an operational metric rather than a rigorous complexity-theoretic quantity. Nevertheless, it provides a useful framework for comparing different quantum algorithms, hardware platforms, and noise conditions from the perspective of computational-information recovery.

\subsection{Recoverability versus Fidelity}

Traditional analyses of noisy quantum computation typically emphasize quantities such as state fidelity, process fidelity, logical error rates, and gate fidelities. These metrics are indispensable for characterizing quantum hardware, but they do not directly measure whether the computational information required by a particular application remains recoverable. Recoverability and fidelity are closely related but conceptually distinct. A computation may exhibit relatively low state fidelity while still preserving the computational information required by the application. Conversely, a computation with high state fidelity may nevertheless fail to produce useful results if the desired computational information cannot be recovered efficiently.

This distinction is particularly important for applications such as quantum machine learning, optimization, parameter estimation, and quantum sensing, where the final objective is often a decision, an optimized parameter, or a physical observable rather than complete reconstruction of the quantum state. From the perspective of RQC, fidelity measures the quality of the quantum evolution, whereas recoverability measures the usefulness of the computational information that survives.

\subsection{Toward Application-Specific Recoverability}

The metrics introduced above are intentionally general. Different classes of quantum algorithms encode computational information in different ways and are therefore expected to exhibit distinct recoverability characteristics. For example, quantum Fourier-transform algorithms encode computational information in spectral features whose visibility gradually degrades under hardware noise. Quantum machine-learning algorithms encode computational information in output probability distributions whose dominant decision labels may remain stable despite moderate perturbations. Quantum sensing and metrology routinely recover physical parameters through repeated measurements and statistical estimation.

Recoverability should therefore not be viewed as a single universal quantity. Rather, it represents a family of application-dependent measures describing how efficiently different forms of computational information can be recovered from imperfect quantum computations. The next two sections illustrate this general framework through two representative examples. Section~\ref{sec:QFT} considers parameter recovery using the quantum Fourier transform, whereas Section~\ref{sec:QML} examines decision recovery in quantum machine learning. Together, these examples demonstrate two complementary manifestations of RQC.

The operational metrics introduced in this section should be viewed as a practical starting point rather than a complete mathematical theory of recoverability. In this Perspective, our objective is to establish an information-centric conceptual framework for analyzing useful quantum computation in the presence of hardware errors, rather than to develop a rigorous information-theoretic or complexity-theoretic formulation. A predictive mathematical theory of recoverability will likely require precise mathematical definitions of computational information, recovery overhead, and recoverability under general noise models, together with rigorous connections to quantum complexity theory, quantum information theory, and statistical decision theory. Developing such a predictive theory represents an important direction for future research.

\section{Example I: Quantum Fourier Transform and Period Estimation}\label{sec:QFT}

\begin{figure*}[t]
\centering
\includegraphics[width=0.8\textwidth]{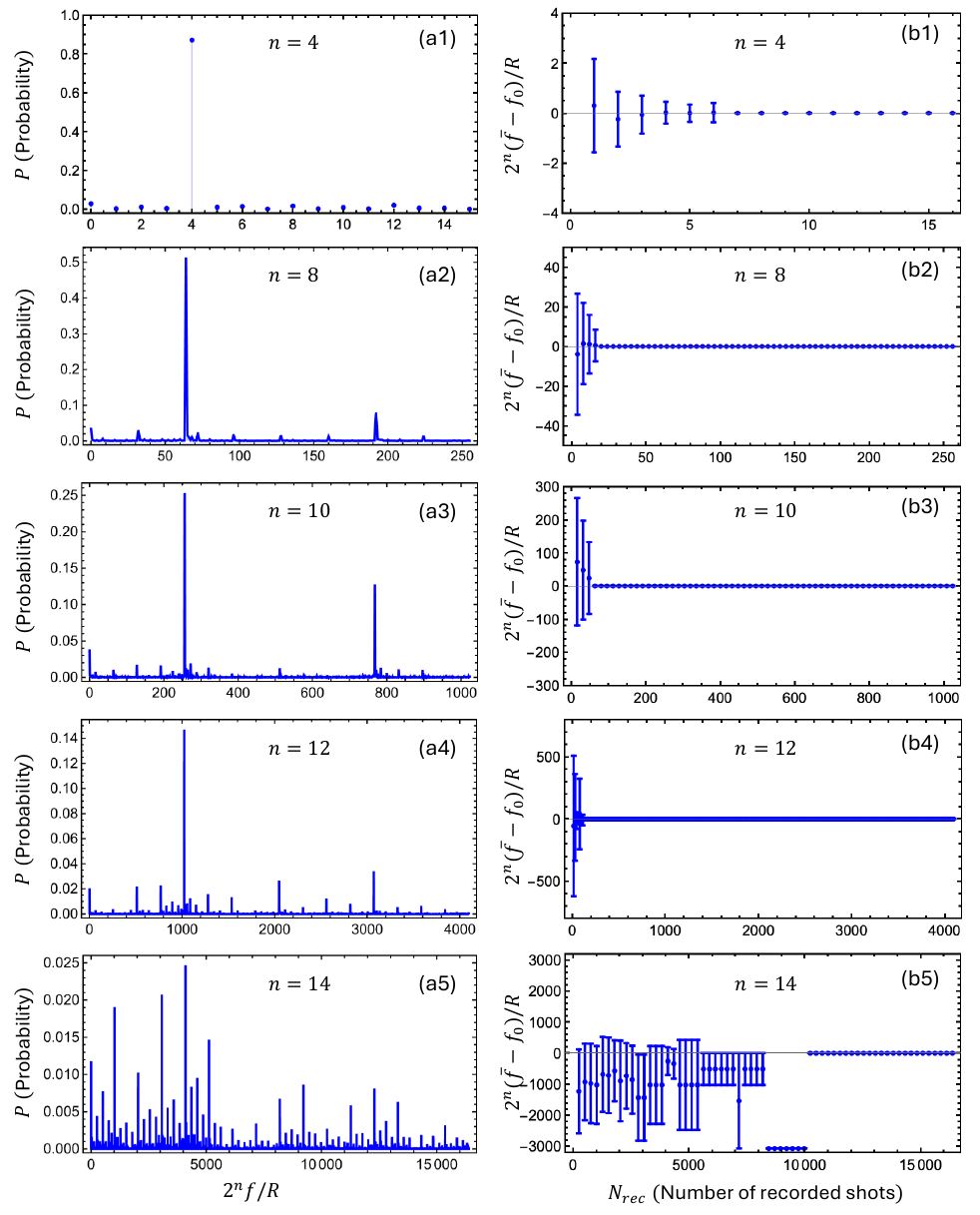}
\caption{
\textbf{Experimental demonstration of RQC using QFT period estimation on IBM quantum hardware.}
(a1)--(a5) Measured probability spectra for $n=4$, $8$, $10$, $12$, and $14$ qubits. The horizontal axis is the normalized frequency coordinate $2^n f/R$, which coincides with the discrete frequency-bin index. The correct signal frequency corresponds to $f_0=R/4$ and is located at $2^{\,n-2}$.
(b1)--(b5) Recovery of the computational information (signal frequency). The vertical axis shows the normalized frequency error,
$2^n(\bar{f}-f_0)/R$,
as a function of the number of recorded shots, $N_{\rm rec}$. Error bars represent one standard deviation obtained from independent subsets of the experimental data. 
}
\label{fig:QFT}
\end{figure*}

One central hypothesis of the RQC framework is that useful computational information may remain recoverable even after the underlying quantum state has been substantially degraded by hardware noise. Quantum Fourier transform (QFT) period estimation provides a natural benchmark for investigating this hypothesis because its computational objective---the estimation of a single frequency---is precisely defined, quantitatively measurable, and theoretically well understood. Moreover, period estimation forms the computational core of several foundational quantum algorithms, including Shor's factoring algorithm and related hidden-subgroup algorithms~\cite{Shor1994,Coppersmith1994}. It therefore provides a natural platform for investigating how quantum noise influences the recoverability of computational information.

For the period-estimation problem considered here, the desired computational information consists solely of the unknown signal frequency. Although the underlying quantum state contains exponentially many probability amplitudes, the computational objective is simply to recover a single frequency. This clear separation between the complexity of the quantum state and the simplicity of the desired output makes QFT period estimation an ideal experimental test of RQC.

We consider a complex signal
\begin{equation}
s(t)=e^{i(2\pi f_0 t+\phi)},
\end{equation}
where $f_0$ is the signal frequency and $\phi$ is the initial phase. Throughout this example we choose
$f_0=R/4$ and $\phi=\pi/10$, where $R$ denotes the sampling rate. In classical signal processing, such as fast Fourier transform (FFT), achieving a frequency resolution $\Delta f=R/(2^n)$ requires $2^n$ sampled data points according to the sampling theorem. Consequently, the classical query complexity scales as $N_{\rm FFT}=2^n$. By contrast, an ideal error-free quantum computer encodes the sampled signal into an $n$-qubit quantum state
\begin{equation}
|\psi_{f_0}\rangle
=
\frac{1}{\sqrt{2^n}}
\sum_{x=0}^{2^n-1}
e^{i(\pi x/2+\phi)}
|x\rangle,
\end{equation}
and performs a QFT. The correct frequency bin, $m_0=2^n f_0/R=2^{\,n-2}$, is then obtained with unit probability in a single execution,
$N_{\rm QFT}=1$.

The information-centric perspective developed in the previous sections naturally leads to the central question of this experiment:

\begin{quote}
\emph{Can the desired frequency remain recoverable when the QFT is implemented on a realistic noisy quantum processor?}
\end{quote}
Hardware noise inevitably broadens the Fourier spectrum and reduces the probability associated with the correct frequency bin. From the perspective of RQC, however, this degradation is not the primary issue. The key question is whether the desired computational information—the signal frequency—can still be recovered with a measurement overhead that preserves a practical quantum advantage.

To investigate this question, we implemented QFT circuits on IBM superconducting quantum processors for $n=4$, $8$, $10$, $12$, $14$, and $16$ qubits. For each circuit size, the target frequency corresponds to the Fourier peak located at $m_0=2^n f_0/R=2^{\,n-2}$. Figure~\ref{fig:QFT} summarizes the experimental results. Panels (a1)--(a5) present the measured probability spectra obtained directly from IBM quantum hardware. The horizontal axis is expressed in the normalized frequency coordinate $2^n f/R$, which coincides with the discrete frequency-bin index. Because the input signal satisfies $f_0=R/4$, the correct frequency always appears at the position $2^{\,n-2}$.

As the circuit size increases, the measured probability spectra evolve in a systematic and physically intuitive manner. Figure~\ref{fig:QFT}(a1)--(a5) show that the dominant Fourier peak remains correctly located at the target frequency for $n=4$ through $n=14$. At the same time, however, the peak probability decreases steadily while an increasingly broad background of competing peaks emerges. These observations are consistent with the cumulative effects of hardware noise during quantum computation.

As the number of qubits and quantum gates increases, accumulated gate errors redistribute probability amplitude from the target Fourier mode into many other computational basis states. Because the dimension of the Hilbert space grows exponentially with the number of qubits, the number of accessible error states also increases rapidly. The experimental consequence is a progressive broadening of the Fourier spectrum accompanied by a steady reduction of the probability associated with the correct frequency. The measured peak probability decreases from approximately $0.87$ for $n=4$ to $0.51$ for $n=8$, $0.25$ for $n=10$, $0.15$ for $n=12$, and only $0.025$ for $n=14$. By $n=16$ (not shown in Fig.~\ref{fig:QFT}), the correct peak is no longer distinguishable from competing peaks generated by accumulated hardware errors, indicating that the QFT spectrum itself has become severely degraded.

Viewed solely from the perspective of quantum-state fidelity, these observations might appear discouraging. The central question posed by RQC, however, is fundamentally different. Rather than asking whether the probability spectrum remains nearly ideal, we ask whether the desired computational information—the signal frequency—can still be recovered despite the degradation of the underlying quantum state. To answer this question, we analyze the complete measurement record rather than individual probability spectra. The experimental data are divided into independent groups containing $N_{\rm rec}$ repeated quantum executions. For each group, the recovered frequency is defined as the frequency bin with the largest number of recorded counts, $\hat{f} = \arg\max_k N_k$, where $N_k$ denotes the number of occurrences measured in the $k$th frequency bin. Repeating this procedure for all independent groups yields a distribution of recovered frequencies, from which the mean recovered frequency $\bar{f}$ and its standard deviation $\sigma_f$ are determined. Figures~\ref{fig:QFT}(b1)--(b5) plot the normalized frequency error, $2^n(\bar{f}-f_0)/R$, as a function of the number of recorded shots. The error bars represent one standard deviation obtained from independent subsets of the experimental data.

Several important observations immediately emerge. Although the probability spectra shown in Fig.~\ref{fig:QFT}(a) become progressively broader with increasing circuit size, the recovered frequency converges steadily toward the correct value as additional quantum executions are accumulated. Even after the probability associated with the correct Fourier peak has been reduced by nearly two orders of magnitude, repeated measurements together with straightforward classical post-processing remain sufficient to recover the desired frequency with high accuracy over a broad range of circuit sizes. This behavior highlights the central distinction between quantum-state fidelity and computational-information recoverability. The degradation of the quantum state does not immediately imply the loss of the computational information. Instead, useful information may remain recoverable long after substantial distortion has appeared in the measured probability spectrum.

Because one frequency bin corresponds to the frequency resolution achieved by a classical FFT using $2^n$ samples, we define the recoverability threshold, $N_{\rm rec}^{*}$, as the minimum number of repeated quantum executions required for the frequency uncertainty to satisfy $2^n\sigma_f/R<1$. This criterion places noisy QFT, ideal QFT, and classical FFT on a common footing by requiring all three approaches to achieve the same effective frequency resolution. The recoverability efficiency introduced in Sec.~\ref{sec:Metrics} therefore becomes
\begin{equation}
\eta_{\rm QFT}
\simeq
\frac{2^n}{N_{\rm rec}^{*}},
\end{equation}
which compares the number of classical samples required by the FFT with the number of repeated noisy-QFT executions required to recover the same computational information. The experimentally determined recoverability thresholds and efficiencies are summarized in Table~\ref{tab:QFT}.

\begin{table}[t]
\caption{\textbf{Recoverability threshold $N_{\rm rec}^{*}$ and recoverability efficiency $\eta_{\rm QFT}$ for QFT period estimation on IBM quantum hardware.}
}
\label{tab:QFT}
\begin{ruledtabular}
\begin{tabular}{ccc}
$n$ & $N_{\rm rec}^{*}$ & $\eta_{\rm QFT}$ \\
\hline
4  & 3      & 5.33 \\
8  & 19     & 13.47 \\
10 & 80     & 12.80 \\
12 & 121    & 33.85 \\
14 & 10200  & 1.61 \\
16 & Failure & 0 \\
\end{tabular}
\end{ruledtabular}
\end{table}

Several important conclusions can be drawn from these results. First, a substantial recoverable quantum advantage is observed for $n=4$--$12$. Although the measured probability spectra are progressively distorted by hardware noise, the desired computational information—the signal frequency—remains recoverable using far fewer repeated quantum executions than the $2^n$ samples required by the classical FFT. Throughout this regime, $\eta_{\rm QFT}\gg1$, demonstrating that useful computational information survives despite significant degradation of the underlying quantum state.

Second, a pronounced transition occurs at $n=14$. Although the dominant Fourier peak remains correctly located, several competing peaks become comparable in amplitude. Consequently, the number of repeated executions required for reliable recovery increases dramatically to approximately $10^4$, reducing the recoverability efficiency to $\eta_{\rm QFT}\approx1.6$. This behavior marks the onset of a transition between a regime in which computational information remains efficiently recoverable and one in which the recovery overhead nearly eliminates the practical quantum advantage.

Finally, at $n=16$, the correct frequency can no longer be recovered reliably, even after extensive repeated measurements. Additional quantum executions no longer improve the recovered result because the desired Fourier peak has become indistinguishable from competing peaks generated by accumulated coherent and stochastic errors. At this point, the computational information itself has become unrecoverable.

These observations naturally divide the experimental results into three computational regimes:
\begin{enumerate}
\item \textbf{Recoverable regime ($n=4$--$12$):}
Useful computational information remains recoverable with substantial quantum advantage.

\item \textbf{Marginal regime ($n=14$):}
The desired information remains recoverable, but the recovery overhead nearly exhausts the available quantum advantage.

\item \textbf{Unrecoverable regime ($n=16$):}
The computational information can no longer be recovered reliably from the noisy quantum outputs.
\end{enumerate}
Taken together, these observations provide experimental support for the central concepts underlying the RQC framework in the specific context of QFT-based period estimation. Substantial degradation of the quantum state does not necessarily imply the loss of useful computational information. Instead, the practical usefulness of noisy quantum computation is ultimately determined by whether the desired information remains recoverable with an overhead that preserves quantum advantage.

The QFT experiment illustrates a clear distinction between quantum-state fidelity and computational-information recoverability. The former characterizes the quality of the quantum evolution, whereas the latter determines whether the computation remains practically useful. This distinction forms the conceptual foundation of RQC and motivates the broader question of how recoverability manifests across different classes of quantum algorithms.

The following section explores whether similar information-centric ideas may also apply to a fundamentally different class of quantum algorithms. Unlike QFT period estimation, where the desired output is a continuous physical parameter, quantum machine learning seeks a discrete computational decision. Despite these fundamentally different computational objectives, both examples reveal the same underlying principle: useful quantum computation depends on the recoverability of task-specific information rather than on perfect preservation of the complete quantum state.


\section{Example II: Recoverability in Quantum Machine Learning}\label{sec:QML}

The QFT example presented in the previous section demonstrates that computational information in the form of a physical parameter can remain recoverable despite substantial degradation of the underlying quantum state. RQC, however, is not restricted to parameter-estimation problems. Many quantum algorithms ultimately seek a computational decision rather than a continuous physical parameter. Quantum machine learning (QML) provides a representative example of this broader class of applications.

In supervised quantum classification, the objective is not to reconstruct the complete quantum state but simply to determine the correct output label. From the perspective of RQC, this distinction is significant. The desired computational information is the classification decision itself, rather than the exponentially large collection of probability amplitudes describing the underlying quantum state. Consequently, preserving the decision may require substantially less information than preserving the entire quantum state.

This observation suggests that useful quantum classification may remain possible even when substantial errors accumulate during quantum computation~\cite{Biamonte2017,Havlicek2019,Schuld2019}. Although the detailed mechanisms differ from those of QFT period estimation, both problems share the same information-centric viewpoint: the objective is to recover the task-specific information required by the application rather than the complete quantum state.

\subsection{Decision Recovery versus State Recovery}

The distinction between state recovery and decision recovery becomes particularly clear in QML. Consider an $n$-qubit quantum processor described by
\begin{equation}
|\Psi\rangle
=
\sum_{k=0}^{2^n-1}
c_k|k\rangle,
\end{equation}
where the quantum state contains $2^n$ complex probability amplitudes. FTQC seeks to preserve the complete set of amplitudes $\{c_k\}$ throughout the computation by means of active quantum error correction. In contrast, many machine-learning algorithms require only a small amount of information extracted from this exponentially large Hilbert space. For a classification problem involving $C$ classes, the desired output is simply
\begin{equation}
\hat{y}\in\{1,2,\cdots,C\},
\end{equation}
rather than the complete quantum state.

RQC therefore asks a different question. Instead of asking whether the quantum state has been faithfully preserved, it asks whether the computational decision encoded in that state remains recoverable. Recovering the correct decision may require substantially less information than reconstructing the complete quantum state, opening the possibility that useful quantum classification survives even in the presence of considerable hardware noise.

\subsection{A Minimal Quantum Classifier}

To illustrate this idea, consider a minimal quantum classifier consisting of three stages: quantum feature encoding, quantum evolution, and measurement.

An input feature vector $\mathbf{x}$ is first encoded into an $n$-qubit quantum state,
\begin{equation}
|\psi(\mathbf{x})\rangle
=
U_{\rm enc}(\mathbf{x})
|0\rangle^{\otimes n},
\end{equation}
where $U_{\rm enc}$ denotes the feature-encoding circuit. The encoded state subsequently evolves according to
\begin{equation}
|\phi(\mathbf{x})\rangle
=
U
|\psi(\mathbf{x})\rangle,
\end{equation}
where $U$ may represent either a fixed quantum circuit or a trained variational quantum model.

The probability associated with class $i$ is obtained from the final measurement,
\begin{equation}
P_i(\mathbf{x})
=
\sum_{z\in C_i}
|\langle z|\phi(\mathbf{x})\rangle|^2,
\end{equation}
where $C_i$ denotes the subset of computational basis states assigned to class $i$. The predicted class is therefore
\begin{equation}
\hat{y}
=
\arg\max_i P_i.
\end{equation}
Unlike FTQC, successful quantum classification does not require preservation of every probability amplitude. It requires only that the correct decision remain identifiable from the measured probability distribution.

\subsection{Interference-Dominated Decision Making}

One possible physical interpretation of recoverability becomes particularly transparent when the classification probabilities are interpreted in terms of quantum interference. The probability amplitude associated with class $i$ may be expressed as
\begin{equation}
A_i(\mathbf{x})
=
\sum_P
a_i(P;\mathbf{x}),
\end{equation}
where $P$ labels the different computational pathways contributing coherently to the final outcome. The corresponding class probability is
\begin{equation}
P_i(\mathbf{x})
=
|A_i(\mathbf{x})|^2.
\end{equation}
For the correct class, constructive interference enhances the total probability amplitude, whereas destructive interference suppresses competing classes. We refer to this process as \emph{decision interference}, emphasizing that the interference determines the computational decision rather than the complete quantum state.

On a noisy quantum processor, each computational pathway is perturbed,
\begin{equation}
a_i(P;\mathbf{x})
\rightarrow
a_i(P;\mathbf{x})
+\delta_i(P;\mathbf{x}),
\end{equation}
leading to a redistribution of probability among different measurement outcomes. Nevertheless, the correct decision may remain recoverable if the constructive interference responsible for the correct classification continues to dominate the accumulated perturbations. This observation motivates the following working principle.

\medskip

\noindent
\textbf{Interference Dominance Principle.}

\emph{A noisy quantum computation remains recoverable when the constructive quantum interference encoding the desired computational information dominates the accumulated error amplitudes generated by hardware imperfections.}

\medskip

Although qualitative, this principle suggests a possible physical interpretation of RQC. Rather than preserving every detail of the quantum state, useful computation requires only that the dominant interference patterns carrying the task-specific information survive the effects of hardware noise.

\subsection{Recoverability of Classification Decisions}

Because individual quantum measurements are probabilistic, repeated executions may be used to recover the correct classification decision. Suppose that $N_{\rm rec}$ repeated measurements produce $N_i$ occurrences of class $i$. The recovered decision is then
\begin{equation}
\hat{y}
=
\arg\max_i N_i.
\end{equation}
This recovery procedure is directly analogous to the frequency recovery introduced in Sec.~\ref{sec:QFT}. In the QFT example, repeated measurements identify the dominant Fourier component. Here, repeated measurements identify the dominant classification outcome.

Let $N_{\rm rec}^{*}$ denote the minimum number of repeated measurements required to recover the correct classification with a specified confidence level. Following the definition introduced in Sec.~\ref{sec:Metrics}, a recoverability efficiency may be written as
\begin{equation}
\eta_{\rm QML}
=
\frac{T_C}
{N_{\rm rec}^{*}T_Q},
\end{equation}
where $T_Q$ denotes the runtime of a single quantum execution and $T_C$ the runtime of the best available classical algorithm.

Although a detailed quantitative analysis lies beyond the scope of this Perspective, the conceptual behavior closely parallels that observed in QFT period estimation. When decision interference remains dominant, the correct classification can be recovered with only modest measurement overhead. As hardware noise increases, progressively more repeated executions are required, producing a marginal recoverability regime. Beyond a critical level of degradation, the constructive interference encoding the correct decision is overwhelmed by accumulated errors, and the computational information itself becomes unrecoverable.

Together, the QFT and QML examples illustrate two complementary manifestations of RQC. In the former, the desired information is a physical parameter; in the latter, it is a computational decision. Despite their different algorithmic structures, both examples demonstrate the same underlying principle: useful quantum computation depends on the recoverability of task-specific information rather than on perfect preservation of the complete quantum state. A detailed theoretical and numerical investigation of recoverable QML will be presented elsewhere.


\section{A Preliminary Classification of Quantum Applications by Recoverability}\label{sec:Classification}

The examples presented in the previous two sections suggest that recoverability is not a property of a particular quantum algorithm, but rather of the computational information that the algorithm ultimately seeks to extract. Quantum algorithms differ substantially in both the amount and the structure of the information that must be preserved during computation. Some seek only a single physical parameter or a discrete computational decision, whereas others require access to a much larger fraction of the quantum-state information. These differences may lead to markedly different degrees of recoverability.

\begin{figure*}[t]
\centering
\includegraphics[width=0.9\textwidth]{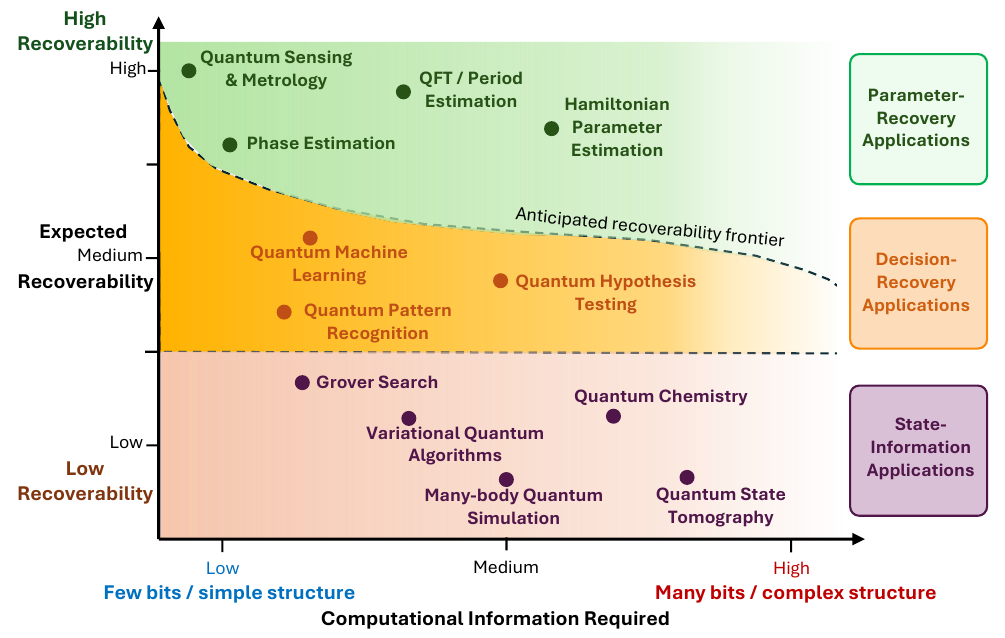}
\caption{
\textbf{Conceptual recoverability landscape of quantum applications.}
Recoverability is expected to decrease as the amount and complexity of the
computational information to be recovered from the quantum state increase.
Parameter-recovery and decision-recovery applications are anticipated to
exhibit higher recoverability than applications requiring preservation of large
amounts of quantum-state information. The placement of individual applications
is illustrative and conceptual rather than quantitative.
}
\label{fig:recoverability_landscape}
\end{figure*}

This observation motivates a preliminary classification of quantum applications according to their expected recoverability. The classification proposed here is intended not as a definitive taxonomy, but as a conceptual framework for organizing future theoretical and experimental investigations of RQC, as shown in Fig.~\ref{fig:recoverability_landscape}.

\subsection{Parameter-Recovery Applications}

Many quantum algorithms ultimately seek to recover one or a few continuous physical parameters. Representative examples include QFT-based period estimation, quantum phase estimation, quantum sensing, quantum metrology, and Hamiltonian parameter estimation~\cite{Shor1994,Coppersmith1994}.

The QFT example presented in Sec.~\ref{sec:QFT} illustrates this class of applications. Although hardware noise substantially broadens the measured Fourier spectrum and reduces the peak probability, the desired information—the signal frequency—remains recoverable over a broad range of circuit sizes through repeated quantum executions and statistical inference. Similarly, quantum sensing and quantum metrology routinely recover physical parameters from noisy measurements through repeated observations and parameter estimation.

Because the desired output consists of only a small number of physical parameters, we expect parameter-recovery applications to exhibit relatively high recoverability. Additional quantum executions can compensate for hardware imperfections while preserving practical quantum advantage, provided that the associated recovery overhead remains sufficiently small.

\subsection{Decision-Recovery Applications}

A second class of quantum algorithms seeks discrete computational decisions rather than continuous physical parameters. Representative examples include QML, quantum pattern recognition, quantum hypothesis testing, and related quantum decision-making algorithms~\cite{Biamonte2017,Havlicek2019,Schuld2019}.

As discussed in Sec.~\ref{sec:QML}, quantum classifiers ultimately seek only the predicted decision label rather than the complete quantum state. Consequently, the amount of information that must be recovered is typically much smaller than the exponentially large information encoded in the Hilbert space. Faithful reconstruction of every quantum amplitude is therefore unnecessary for successful computation. The Interference Dominance Principle introduced in Sec.~\ref{sec:QML} provides one possible physical mechanism underlying recoverability in these applications. When the constructive quantum interference responsible for the correct decision remains dominant over accumulated hardware errors, the desired computational information may remain recoverable even though the underlying quantum state has undergone substantial degradation.

Although detailed theoretical and numerical investigations remain to be carried out, decision-recovery applications may likewise exhibit relatively high recoverability because only a limited amount of task-specific information must ultimately be extracted from the quantum processor.

\subsection{State-Information Applications}

A third class of applications requires substantially richer information from the quantum state itself. Representative examples include quantum chemistry, general quantum simulation, many-body quantum dynamics, and quantum state tomography~\cite{McArdle2020}. These applications often require recovery of a large collection of observables, correlation functions, or, in the limiting case of quantum state tomography, reconstruction of the complete quantum state.

Because these applications require preservation of a much larger fraction of the information encoded in the quantum state, hardware errors are expected to have a more direct impact on the final computational outcome. Although repeated measurements remain valuable for reducing statistical uncertainty, they cannot recover quantum information that has already been irreversibly lost through decoherence or other hardware imperfections.

Recoverability in these applications may therefore be more limited, and FTQC may ultimately remain indispensable for large-scale implementations. Nevertheless, identifying which observables remain recoverable despite imperfect quantum evolution remains an important open question deserving future investigation.

\subsection{Recoverability as an Organizing Principle}

We hypothesize that recoverability depends primarily on the nature and structure of the computational information that must ultimately be preserved rather than on the specific quantum algorithm itself. Figure~\ref{fig:recoverability_landscape} summarizes this perspective conceptually. Applications whose outputs consist primarily of a few physical parameters or discrete computational decisions are generally expected to exhibit higher recoverability than applications requiring preservation of a large fraction of the quantum-state information. The locations of individual applications in Fig.~\ref{fig:recoverability_landscape} are qualitative and are intended only to illustrate the proposed conceptual framework.

These observations motivate the following working hypothesis.

\medskip

\noindent
\textbf{Recoverability Hypothesis.}

\emph{We hypothesize that the recoverability of a quantum computation depends primarily on the amount and structure of the computational information that must be preserved, rather than solely on the complexity of the underlying quantum state.}

\medskip
This hypothesis does not imply that recoverability depends solely on the amount of information. Circuit depth, hardware noise, gate fidelity, measurement errors, algorithmic structure, and classical post-processing all influence recoverability. Rather, it suggests that the computational objective itself plays a central role in determining whether useful computational information survives imperfect quantum execution.

The classification proposed here is intentionally preliminary. Many important quantum algorithms—including Grover search, variational quantum algorithms, large-scale quantum chemistry simulations, quantum communication, distributed quantum computing, and quantum networking—may occupy intermediate regions of the recoverability landscape whose properties remain largely unexplored. Developing quantitative criteria for locating quantum algorithms within this landscape represents one of the central challenges for future research.

\section{Research Roadmap}\label{sec:Roadmap}

RQC provides an information-centric framework for understanding useful quantum computation in the presence of hardware imperfections. The preceding sections suggest that useful computational information may remain recoverable across a broad intermediate regime between today's noisy quantum processors and tomorrow's fully fault-tolerant quantum computers. Transforming this conceptual framework into a predictive theory, however, will require substantial advances in both theory and experiment. Rather than presenting an exhaustive list of open problems, we highlight several research directions that we believe will be particularly important for establishing RQC as a mature area of quantum information science.

\subsection{Toward a Predictive Theory of Recoverability}

At present, recoverability is characterized primarily through operational quantities such as recovery overhead and recoverability efficiency. An important theoretical challenge is to develop a predictive framework capable of relating recoverability directly to quantum circuit structure, hardware noise, and the nature of the computational information required by different quantum algorithms.

One particularly important question is whether recoverability can be estimated analytically before a quantum computation is performed. Such a capability would provide valuable guidance for quantum algorithm design, hardware development, and resource estimation. More fundamentally, it could reveal the conditions under which computational information remains recoverable despite imperfect quantum evolution and identify the boundaries separating recoverable and unrecoverable quantum computation.

\subsection{Recoverability-Aware Quantum Algorithms}

Most existing quantum algorithms are designed under the implicit assumption that maximizing quantum-state fidelity is the primary objective. RQC suggests an alternative design philosophy: optimize the preservation of computational information rather than the complete quantum state. This perspective opens the possibility of developing quantum algorithms that explicitly maximize recoverability by enhancing the robustness of task-specific information against realistic hardware imperfections. Likewise, quantum compilers, circuit optimization strategies, and error-mitigation techniques may be designed to preserve the dominant computational interference pathways rather than simply minimizing conventional hardware-error metrics. This direction appears particularly promising for variational quantum algorithms and other hybrid quantum-classical approaches, whose performance depends on the interplay among ansatz design, hardware noise, measurement overhead, and classical optimization~\cite{Cerezo2021}.

\subsection{Recoverability-Oriented Hardware Benchmarks}

Current benchmarks of quantum processors primarily emphasize hardware-oriented metrics such as gate fidelity, logical error rates, and quantum volume~\cite{Acharya2023,Cross2019}. Although indispensable for characterizing quantum hardware, these quantities do not directly measure the practical usefulness of noisy quantum computation for a specific computational task.

Recoverability motivates a complementary class of application-oriented benchmarks. Rather than asking how accurately a quantum processor preserves a quantum state, one may instead ask how efficiently the computational information required by a given application can be recovered under realistic operating conditions. Such benchmarks could enable more meaningful comparisons among different quantum hardware platforms from the perspective of practical quantum applications.

\subsection{Exploring the Recoverability Landscape}

The examples presented in this Perspective represent only the first steps toward understanding RQC. Many important quantum algorithms and applications—including Grover search, variational quantum optimization, quantum chemistry, quantum simulation, quantum communication, distributed quantum computing, quantum networking, and quantum sensing~\cite{Grover1996,McArdle2020,Degen2017}—remain to be investigated from the perspective of recoverability.

An especially important direction will be to understand how recoverability depends on the type of computational information being extracted. Developing quantitative recoverability maps for different classes of quantum algorithms could ultimately lead to a predictive understanding of which applications can benefit from noisy quantum hardware and which fundamentally require fault-tolerant operation.

More broadly, RQC should not be viewed as an alternative to FTQC, but as a complementary paradigm occupying the broad intermediate regime between today's noisy quantum processors and tomorrow's fully fault-tolerant quantum computers. Understanding this intermediate regime may prove to be as important as the pursuit of fault tolerance itself, because it determines when useful quantum advantage can be achieved on realistic quantum hardware.

\section{Conclusions}\label{sec:Conclusions}

FTQC provides the ultimate route toward scalable, arbitrarily accurate quantum information processing through quantum error correction. This vision remains indispensable for realizing large-scale quantum computers. At the same time, the substantial hardware overhead required for fault tolerance raises a fundamental question: must useful quantum computation wait until fully fault-tolerant quantum hardware becomes available?

In this Perspective, we have proposed RQC as an information-centric paradigm for quantum computing with errors. Rather than asking whether a quantum computation faithfully preserves the complete quantum state, RQC asks whether the computational information required to accomplish a given task remains recoverable despite imperfect quantum execution. This shift in perspective distinguishes RQC from existing approaches centered on quantum error correction, quantum error mitigation, or hardware-oriented performance metrics. The central question is no longer whether quantum evolution is perfect, but whether useful computational information can be recovered efficiently while preserving quantum advantage.

To support this perspective, we introduced recoverability as an operational criterion together with practical metrics based on recovery overhead and recoverability efficiency. The QFT period-estimation experiment provided a representative demonstration of how task-specific computational information can remain recoverable despite substantial degradation of the underlying quantum state. QML offered a complementary conceptual example illustrating how computational decisions may likewise remain recoverable through the preservation of dominant quantum interference patterns. Together, these examples suggest that recoverability is not unique to a particular quantum algorithm, but instead reflects a broader property of quantum computation in the presence of noise.

More broadly, we proposed a preliminary classification of quantum applications according to the type of computational information they seek to recover. We expect applications involving parameter recovery or decision recovery to exhibit substantially higher recoverability than applications requiring preservation of a large fraction of the quantum-state information. This perspective naturally motivates a broader research program encompassing predictive theories of recoverability, recoverability-aware quantum algorithms, application-oriented hardware benchmarks, and quantitative recoverability maps for different classes of quantum algorithms.

RQC does not replace FTQC, nor does it diminish the importance of quantum error correction. Rather, the two paradigms address complementary objectives. FTQC seeks to preserve the complete logical quantum state, whereas RQC provides a complementary framework for evaluating whether computational information remains recoverable despite imperfect quantum execution. Together, they may provide a broader framework for understanding useful quantum computation across the full spectrum of noisy and fault-tolerant quantum hardware.

Many important questions remain open. The ideas introduced in this Perspective are intended not as a complete theory, but as the starting point for a broader research program. Although only one representative application has been experimentally examined here, we anticipate that the proposed framework will motivate further theoretical and experimental investigations across diverse classes of quantum algorithms and help define the emerging science of recoverability in quantum computation. Ultimately, the future of quantum computing may depend not only on how well we preserve quantum states, but also on how effectively we preserve the computational information required to solve meaningful problems.

\begin{acknowledgments}
We acknowledge the use of IBM Quantum services for this work. The views expressed are those of the authors and do not reflect the official policy or position of IBM or the IBM Quantum team. The quantum circuits were executed using Qiskit v2.3.0 on the ibm\_kingston processor. The work was supported by NSF (Grant
Nos. 2503230 and 2228725).
\end{acknowledgments}

\section*{Data Availability}
The data that support the findings of this article are openly available~\cite{RQC_2026_Zenodo}.

\section*{Conflict of interest}
The author has no conflict to disclose.

\bibliography{references}

\end{document}